# A Matching Game for Data Trading in Operator-Supervised User-Provided Networks

Beatriz Lorenzo and F. Javier Gonzalez-Castano
Telematics Department, University of Vigo, Spain
Emails: {blorenzo, javier}@gti.uvigo.es

*Abstract*—In this paper, we consider a recent cellular network connection paradigm, known as a user-provided network (UPN), where users share connectivity and act as an access point for other users. To incentivize user participation in this network, we allow the users to trade their data plan and obtain a profit by selling and buying leftover data capacities (caps) from each other. We formulate the data trading association between buyers and sellers as a matching game. In this game, buyers and sellers rank each other based on preference functions that capture the buyers' demand for data and QoS requirements, the data available for purchase from the sellers and energy resources. We show that these preferences are interdependent and influenced by existing network-wide matching. For this reason, the game can be classified as a one-to-many matching game with externalities. To solve the game, we propose a distributed algorithm that combines notions from matching theory and market equilibrium. The algorithm enables the players to self-organize into a stable matching and ensures dynamic adaptation of price to data demand and supply. The properties of the resulting matching are discussed. We also calculate operator gains and the benchmark price that will encourage users to join the UPN. Simulation results show that the proposed algorithm yields average utility per user improvements of up to 25% and 50% relative to random matching and worst case utility, respectively.

## I. INTRODUCTION

The increasing demand for mobile data in current cellular networks and the proliferation of advanced handheld devices place user-provided networks (UPNs) in a prominent position for next-generation network architectures [1], [2]. In UPN, users share their connectivity and provide Internet connection for others without additional network infrastructure costs. Some UPN services rely on fixed access points like FON [3], while others are more flexible, and rely on mobile devices such as smartphones and portable devices [1], [4]. Several UPN business models have recently been implemented by different startups and operators [4], [5]. Open Garden [4], for instance, enables mobile users to create a mesh network and share their Internet connections without the intervention of a network operator, while Karma [5], virtual mobile operator, enables its subscribers to act as mobile WiFi hotspots (MiFi) to serve non-subscribers by offering in return free data. The adoption of UPNs by major network operators emphasizes the potential of these networks to generate gains for both users and operators. However, their success heavily depends on users' willingness to contribute their resources.

In this paper, we consider an operator-supervised UPN where users share their mobile connection and act as access points for users in their vicinity. Motivated by the recently launched traded data plans [6], where wireless service providers (WSP) allow users to sell and buy leftover data capacities from each other, we incorporate the concept of data plan trading into the UPN as an incentive for users to participate in this network. We study and design novel strategies for buyer-seller data trading associations, data trading price, and pricing mechanisms that will encourage users to join the UPN. In our model, we broaden the use of UPNs to the following cases: a) if the operator is not able to satisfy the users' QoS requirements, he will encourage them to transmit through the UPN in return for compensation; b) if users use up their data plan, they can buy additional data through the UPN. No additional traffic control measures are needed and the operator will receive a profit proportional to the amount of data traded in the UPN.

Several recent studies proposed incentive mechanisms for UPNs motivated by the commercial practices of Open Garden [7] and Karma [8]. A scheme based on the Nash bargaining solution is presented for an Open Garden-like UPN in [7] to incentivize mobile users to share their connectivity and resources both fairly and efficiently. In [8] the operator determines a free quota reimbursement and data price charged to each user to maximize the seller´s revenue in a Karma-like UPN. The interaction is modelled as a non-cooperative Stackelberg two-stage game. However, these works focus on either a fixed network topology [7] or on the iteration of a single seller and its buyers [8]. In [1], a dynamic network architecture based on the UPN concept is proposed. The authors consider the users' QoS requirements on access point selection and provide a set of contracts based on available connectivity (wired or wireless). However, they focus on centralized social welfare optimization and ignore competition between buyers and sellers.

In this paper, we consider an operator-supervised UPN where users can trade their data plan and earn a profit by selling and buying leftover data capacities from each other. The WSP supervises the trading and ensures that the sellers' trading revenue and the buyers' purchased data are reflected in their bills. We formulate the buyer-seller data trading association as a matching game. In this game, buyers and sellers need to rank one another based on preference functions that capture data demand and QoS requirements in the case of buyers and, available data and energy consumption in the case of sellers. These preferences are independent and strongly influenced by the formation of other buyer-seller links. The proposed game can be classified as a many-to-one matching game with externalities. To solve this game, a distributed algorithm is developed that enables users to self-organize into a stable matching; in addition, the optimum trading price is derived from market equilibrium. In such a data market, we are interested in answering two key questions: a) how do different choices by users affect each other's decision to join the UPN? and, b) how does the price impact that decision?

We assess the performance of the algorithm using simulations and show that it yields significant performance gains for all parties involved.

The rest of the paper is organized as follows. In Section II, we introduce the system model and in Section III we describe the analytical market model for data trading. In Section IV, we develop our matching-theory approach to data trading. In Section V, we present the numerical results and finally, in Section VI we conclude our paper.

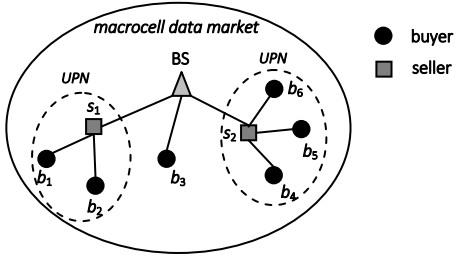

Fig. 1. An example of a macro-cell data market.

## II. SYSTEM MODEL

We consider an uplink transmission in a single macro-cell, as depicted in Fig. 1. It is assumed that a monopolistic wireless service provider (WSP) charges its subscribers a fixed fee for a maximum data volume in a month. The WSP limits excessive data usage by charging overage fees per usage exceeding the monthly data cap. Besides, leftover data cannot be utilized in subsequent months. This hybrid pricing scheme is commonly adopted in the cellular service market to control traffic load [1], [8]-[9]. For a better utilization of resources, the WSP allows its subscribers to sell leftover data caps while acting as an access point for other users in their vicinity. Such a network is referred to as a UPN.

We model the behavior of users in the macro-cell data market as buyers and sellers. Let $\mathcal{B} = \{1,\ldots, B\}$ and $\mathcal{R} = BS \cup \{1,\ldots,S\} = \{0,\ldots,S\}$ denote the set of all buyers and sellers, respectively. The bandwidth allocated to buyer $b$ by seller $s$ is denoted by $w_{bs}$, and the buyer's transmission power is $P_b$. We consider that sellers use a time-division multiple-access scheduler in which each time slot has a duration $\tau_t$. Using this uplink transmission system implies that users assigned to the same seller do not interfere with each other, i.e., there is no intra-access point interference. Note that the analysis undertaken in this paper is equally applicable to other multiple access and scheduling schemes [10]-[12]. Further, as the UPN is enabled by users and interference is difficult to predict, we consider that the operator assigns separate macro-cell channels for the UPN used. In this regard, buyers transmitting in adjacent UPNs will interfere with each other.

Then, the capacity of the link between buyer $b$ and the BS ($s = 0$) is,

$$c_{b0}(\gamma_{b0}) = w_{b0}\log(1+\gamma_{b0}), \quad (1)$$

where $\gamma_{b0} = P_b h_{b0}/\sigma^2$ is the SNR, with $h_{b0}$ indicating the channel gain between buyer $b$ and BS, and $\sigma^2$ the variance of the Gaussian noise.

Similarly, the capacity between buyer $b$ and seller $s$, $s \neq 0$, is given by

$$c_{bs}(\gamma_{bs}) = w_{bs}\log(1+\gamma_{bs}), \quad \forall s/s \neq 0 \quad (2)$$

where $\gamma_{bs} = P_b h_{bs}/(\sigma^2 + I_{bs})$ is the SINR with $h_{bs}$ indicating the channel gain between buyer $b$ and seller $s$ and $\sigma^2$ the variance of the Gaussian noise. Here, the interference component $I_{bs} = \sum_{b'\neq b} P_{b'} h_{b's}$, $b' \in \mathcal{B}\setminus\{b\}$ relates to transmissions from other buyers $b'$ to their respective sellers using the same sub-channels, and $P_{b'}$ and $h_{b's}$ denote the transmit power and the channel gain between buyer $b'$ and seller $s$ [20].

Buyers need the resources for a specific contractual period. The data volume transmitted on the link between user $b$ and $s$ depends on the capacity of the link and contract duration $\tau_{bs}$ as

$$Q_{bs} = w_{bs}\log_2(1+\gamma_{bs})\tau_{bs}. \quad (3)$$

If we assume that the BS serves the users within a frame of duration $\Delta_t$, the connectivity availability of the BS is

$$\tau_0 = \Delta_t - \delta_0 \quad (4)$$

where $\delta_0 = \sum_{N_0} \tau_{b0}$ and $N_0$ is the number of users connected to the BS, each requesting a connectivity duration $\tau_{b0}$. We consider that $\Delta_t$ is large enough to serve all subscribers.

Energy cost is a critical parameter for the sellers participating in the UPN. The connectivity availability duration for a seller $s$, $s \neq 0$, depends on the remaining battery duration $\tau_b$ and physical availability $\tau_p$. Connectivity availability is thus

$$\tau_s = min(\tau_b - \gamma\delta_s, \tau_p), \quad \forall s/s \neq 0 \quad (5)$$

where battery duration depletes, proportionally to $\gamma$, with the number of users connected to the seller [13].

The aims of our model are to encourage buyers to select the service (UPN or macro-cell BS) that best satisfies their connectivity requirements in order to maximize their utility, and at the same time enable the sellers and the operator to make a profit. The service duration constraint and traffic dynamics are considered to reflect a real network. The WSP acts as a central controller that supervises transactions (e.g., it ensures that data purchases and sales are reflected in users' monthly bills).

## III. ANALYTICAL MARKET MODEL FOR DATA TRADING

Each buyer has certain minimum requirements in terms of channel quality and availability of service duration (i.e., $\gamma_{bs} \geq \gamma_{b,min}$, $\tau_{bs} \geq \tau_{b,min}$) that a seller must satisfy. We define a connectivity parameter to fulfil these minimum requirements,

$$\beta_{bs} = \begin{cases} 1, & \text{if } \gamma_{bs} \geq \gamma_{b,\min} \text{ and } \tau_{bs} \geq \tau_{b,\min} \\ 0, & \text{otherwise} \end{cases}. \quad (6)$$

Since users can purchase different data caps from the WSP, we denote the initial data caps of buyer $b$ and seller $s$ as $Q_b^i$

and $Q_s^i$, respectively. Users do not know exactly how much data they will use over the coming month. We denote as $e$ the probability that a user will exceed his/her monthly cap. In such a case, the user needs to choose between joining the UPN or paying an additional fee to buy extra data directly from the WSP. Besides, a user may decide to join the UPN even if he/she has enough data available $(1 - e)$ but in this case the macro-cell transmission cannot provide the required level of QoS.

We assume that users may be interested in buying or selling a certain volume of data at a given time during the month. Demand for data volume will exist when a potential buyer $b$ perceives profit in trading. This profit can be defined as the difference between the gain from using the data volume and the price paid to the seller for it.

The utility of the user when transmitting to the macro-cell BS ($s = 0$) is defined as

$$U_{b0}(Q_{b0}) = \beta_{b0}[e_b(f(Q_b^i + Q_{b0}) - Q_{b0}p) + (1-e_b)f(Q_b^i)] \quad (7)$$

where the first term is the utility when the user exceeds the initial data plan $Q_b^i$ with probability $e_b$ and purchases additional data volume $Q_{b0}$ at price $p$ (price per unit of data exceeding the data plan). The second term is the utility when the user has data available with probability $(1 - e_b)$. The function $f(\cdot)$ is a non-decreasing concave function with decreasing marginal satisfaction. A common example of this function is the $\alpha$-fair utility function [6]

$$f(Q) = \frac{\theta Q^{1-\alpha}}{1-\alpha},$$

where $\alpha \in [0,1)$ and $\theta > 0$ is a scaling factor. We use $(\theta_b, \alpha_b)$ to denote the parameters for buyers and $(\theta_s, \alpha_s)$ sellers.

The utility of a buyer $b$ when buying traffic volume $Q_{bs}$ from seller $s$, $s \neq 0$, is

$$U_{bs}(Q_{bs}; \pi_{bs}) = \beta_{bs}[e_b(f(Q_b^i + Q_{bs}) - Q_{bs}\pi_{bs}) + \\ (1-e_b)(f(Q_{bs}) + Q_{bs}(r - \pi_{bs}))] \quad (8)$$

where the first term is the utility when the buyer has exceeded his data plan $Q_b^i$ with probability $e_b$ and $\pi$ is the price per unit of data volume traded. The second term is the utility when the user has not exceeded his data plan $(1-e_b)$ and $r$ is the reward from the WSP for transmitting in the UPN.

In response to demand, the BS ($s = 0$) will obtain a utility:

$$U_0(Q_{b0}) = \beta_{b0}(f(Q_b^i + Q_{b0}) + Q_{b0}p) \quad (9)$$

where $Q_{b0}$ is the additional data purchased when the buyer exceeds the data plan. If the buyer has not exceeded the plan then, $Q_{b0} = 0$ and the utility will depend only on the initial data volume.

Similarly, a seller $s$, $s \neq 0$, will be willing to sell his/her leftover traffic volume in return for a profit. This profit can be defined as the gain from serving a number of buyers and the revenue earned from selling the resource. Thus, the utility of a seller when selling traffic volume $Q_s$ is

$$U_s(Q_{sb}; \pi_{sb}) = \beta_{bs}(f(Q_s^i - Q_{sb}) + Q_{sb}(\pi_{sb} - \xi)) \quad (10)$$

where $Q_s^i$ is the monthly data volume purchased, $\pi$ is the price per unit data volume in the secondary data market and $\xi$ is the energy cost for serving as an AP.

An example of the buyer-seller data trading association in the UPN is shown in Fig. 2, where buyer 1 and 2 select the same seller for data trading.

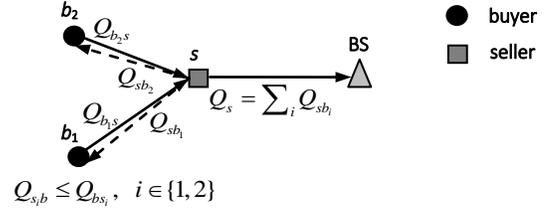

$Q_{s,b} \leq Q_{bs_i}, \quad i \in \{1, 2\}$

Fig. 2. Illustration of data trading in UPN.

By considering the previous definitions, the buyer optimization is as follows

$$\begin{aligned}&\underset{Q_{bs}, \pi_{bs}}{\text{maximize}} \quad U_{bs} \\ &\text{subject to} \quad Q_{bs} \leq \beta_{bs} w_{bs} \log_2(1 + \gamma_{bs})\tau_{bs}\end{aligned} \quad (11)$$

where $\gamma_{bs}$ is the SINR on the link between user $b$ and $s$, $\tau_{bs}$ is the duration of that transmission and $\pi_{bs}$ is the price for the data traded. Buyer $b$ selects the seller who satisfies his/her minimum requirements and the data volume sought is constrained by the capacity of the link.

The seller optimization problem is as follows,

$$\begin{aligned}&\underset{Q_{sb}, \pi_{sb}}{\text{maximize}} \quad \sum_b U_{sb} \\ &\text{subject to} \quad \sum_b \beta_{bs} Q_{sb} \leq Q_s^i, \ \forall s, \ s \neq 0 \\ &\qquad\qquad\ \ Q_{sb} \leq Q_{bs}\end{aligned} \quad (12)$$

where the total data volume sold by seller $s$, $s \neq 0$, $Q_s = \sum_b \beta_{bs} Q_{sb}$, should not exceed his/her data volume.

Solving the buyer-seller association using classical optimization techniques is an NP-hard problem [1], which depends on the number of buyers and sellers in the network. In Section IV, a new *data trading algorithm* is presented to solve buyer and seller optimization in a self-organized distributed manner. The algorithm combines matching theory to form buyer-seller associations and market equilibrium to obtain the trading price. The matching of buyers and sellers results in a trading matrix $\mathbf{T} = [t_{bs}]$ with entries $t_{bs} = 1$, if buyer $b$ buys data from seller $s$, or 0 otherwise. It is worth noting that the trading matrix also defines the topology.

We define the revenue of the WSP in our data market as,

$$U_w = \sum_b [e_b(t_{b0}Q_{b0}p + \sum_{s,s\neq 0}t_{bs}vQ_{bs}\pi_s) - \\ (1-e_b)\sum_{s,s\neq 0}t_{bs}Q_{bs}r] \quad (13)$$

where the first term is the revenue obtained when the buyer exceeds the data plan, $t_{bs}$ is the optimum matching between $b$ and $s$ obtained in Section IV and $v$ is the proportionality coefficient. The second term refers to the reward $r$ that the WSP pays to the buyer as an incentive to join the UPN when it cannot provide the QoS requirements.

## IV. DATA TRADING AS A MATCHING GAME WITH EXTERNALITIES

In this section, the framework of matching games is used to develop a self-organized buyer-seller association algorithm to solve the optimization problem while avoiding combinatorial complexity [14]. We formulate the problem as a one-to-many matching game between buyers and sellers in which each buyer can be associated with only one seller, and sellers can admit a certain quota of users. Buyers and sellers rank one another based on utility functions that capture their preferences. These preferences are interdependent and are influenced by the existing matching. Hence, the proposed game can be classified as a matching game with *externalities*. We formally define the notion of matching:

**Definition 1**. A matching (or association) $\eta$ is a function from the set $\mathcal{B} \cup \mathcal{R}$ into the set $\mathcal{B} \cup \mathcal{R}$ such that,

1) $s = \eta(b)$ if and only if $b = \eta(s)$,
2) $|\mu(b)| = 1$ and $|\mu(s)| \leq n_{ab}$,

where $n_{ab}$ is the number of buyers that can be served by seller $s$ with $Q_s^i$.

A preference relation $\succ_b$ for buyer $b$ over the set of sellers $\mathcal{R}$ is defined as,

$$(s, \eta) \succ_b (s', \eta') \Leftrightarrow U_{bs}(\eta) > U_{bs'}(\eta'). \qquad (14)$$

where sellers $s$, $s' \in \mathcal{R}$, $s \neq s'$, and $s = \eta(b)$, $s' = \eta'(b)$. Analogously, a preference relation $\succ_s$ for seller $s$ over the set of buyers $\mathcal{B}$ is defined as,

$$(b, \eta) \succ_s (b', \eta') \Leftrightarrow U_{sb}(\eta) > U_{sb'}(\eta'). \qquad (15)$$

where buyers $b$, $b' \in \mathcal{B}$, $b \neq b'$ and $b = \eta(s)$, $b' = \eta'(s)$.

### A. Buyer and Seller Preferences

We define the existence of a link between buyer $b$ and seller $s$ by variable $t_{bs;\eta} = \{t_{bs} | \eta, t_{bs} \leq \beta_{bs}\}$ conditioned on the current matching $\eta$ and QoS requirements. The aim of each buyer (seller) is to maximize his/her own utility, or equivalently, to become associated with the most preferred seller (buyer).

The buyer and seller utilities given in (7)-(10) are redefined below to include their connectivity constraints in the utility function by applying the penalty function method with

$$\psi_{bs} = \begin{cases} 0, & Q_{bs} \leq \beta_{bs} w_{bs} \log_2(1 + \gamma_{bs})\tau_{bs} \\ 1, & \text{otherwise} \end{cases} ; \quad \phi_{bs} = \begin{cases} 0, & \sum_b \beta_{bs} Q_{sb} \leq Q_{os} \\ 1, & \text{otherwise} \end{cases}$$

Then, the modified buyer utility is

$$\tilde{U}_b(Q_{bs}; \pi_{bs}, \eta) = t_{bs} U_b(Q_{bs}; \pi_{bs}) - \kappa \psi_{bs} \qquad (16)$$

where $\kappa$ is the penalty factor, $\kappa \geq 0$. If the buyer data demand violates the connectivity constraint, the penalty term $\kappa$ will reduce the data requested.

Similarly, the modified seller utility is

$$\tilde{U}_s(Q_{sb}; \pi_{sb}, \eta) = \sum_b t_{bs} U_s(Q_{sb}; \pi_{sb}) - \rho \phi_{bs} \qquad (17)$$

where $\rho$ is the penalty factor for the seller, $\rho \geq 0$. If the data volume offered by the seller exceeds his/her overall data cap, the penalty term $\rho$ will reduce the data offered.

### B. Proposed solution

Given the formulated *data trading game*, our goal is to find a stable matching, which is one of the key solution concepts of matching theory. Due to these externalities, the traditional solutions based on the deferred acceptance algorithm used in [14] are unsuitable since the ranking of preferences changes as the matching forms. Thus, we look at a new stability concept based on the idea of swap matching [15], [19] and extend it to define *seller-swap* and *service-swap* in our data trading market. Then, the optimum price for the data traded in such a matching scenario is obtained.

**Definition 2**. Given a matching $\eta$, a pair of buyers $b$, $b' \in \mathcal{B}$ and sellers $s$, $s' \in \mathcal{R}$ where $(b, s)$, $(b', s') \in \eta$, a *seller-swap* is defined as $\eta_{ss'}^b = \{\eta \setminus (b, s)\} \cup (b, s')$.

**Definition 3.** A *service-swap* is a seller-swap where seller $s$, $s' \in \mathcal{R}$ with $s$ or $s' = 0$.

A matching is stable if there are no swap matchings (i.e., seller-swap or service-swap) $\eta_{ss'}^b$, such that:

- $\forall x \in \{b, b', s, s'\}$, $U_{x, \eta_{ss'}^b}(\eta) > U_{x, \eta(x)}(\eta)$ and
- $\exists x \in \{b, b', s, s'\}$, $U_{x, \eta_{ss'}^b}(\eta) \geq U_{x, \eta(x)}(\eta)$

A matching $\eta$ is said to be *stable* if there is no buyer $b'$ or seller $s'$, for which $(b', \eta') \succ_s (b, \eta)$, or $(s', \eta') \succ_b (s, \eta)$. The stability is reached by guaranteeing that swaps only occur if all members involved will improve their utilities. And thus, the order of preferences for each player not involved in the swap will be unaltered.

To solve the formulated matching game, we propose a novel distributed algorithm for data trading (Algorithm 1) that enables the players to self-organize into a stable matching that guarantees the required QoS. The proposed algorithm consists of two main stages. Stage 1 focuses on buyer and seller discovery, association and swap-matching evaluation, and Stage 2 determines the data trading price.

First, we assume an initial price for the data $\pi_{bs}(0)$ that will be updated later on based on data demand and supply. The next stage is to form buyer-seller associations comprising a seller and a set of buyers. Each buyer selects a set of sellers that satisfy his/her QoS requirements and sorts them in decreasing order according to their respective utility function in (16). The buyer then selects the top utility-providing sellers, denoted by set $\mathcal{R}_b$. Similarly, each seller $s \in \mathcal{R}_b$ may also be selected by a set of buyers. Using the same selection process, the seller accepts the top utility-providing buyers denoted by $\mathcal{B}_s$ according to their utility in (17). However, because of externalities the order of the buyers' preferences may have changed since the seller selection. Based on the current matching $\eta$, buyers and sellers update their utilities and preferences over one another and perform service-swap and seller-swap until a stable matching is found. Next, we determine the equilibrium price for each UPN by market equilibrium [16].

The data demand and supply functions for each UPN are obtained by differentiating the utilities in (16) and (17) with respect to $Q_{bs}$ and $Q_{sb}$, respectively,

$$\mathcal{D}_b = Q_{bs} = e_b\left[\left(\frac{\theta_b}{\pi_{bs}+\kappa\psi_{bs}}\right)^{1/\alpha_b} - Q_b^i\right] + (1-e_b)\left(\frac{\theta_b}{\pi_{bs}-r+\kappa\psi_{bs}}\right)^{1/\alpha_b} \quad (18)$$

$$\mathcal{S}_s = Q_{sb} = Q_s^i - \left(\frac{\theta_s}{\pi_{bs}-\xi-\rho\phi_{bs}}\right)^{1/\alpha_s} \quad (19)$$

where $(\theta_b, \alpha_b)$ and $(\theta_s, \alpha_s)$ denote the parameters for buyers and sellers in the utility $f(\cdot)$, respectively and $\kappa$ and $\rho$ are the penalty term for the buyer and seller, respectively.

The total data demand and supply for a particular seller at equilibrium is $\sum_{b\in\mathcal{B}_s}\mathcal{D}_b = \sum_{s\in\mathcal{R}_b}\mathcal{S}_s$. Then, the equilibrium price is derived iteratively using the following equation,

$$\pi_{bs}(t+1) = \pi_{bs}(t) + \sigma_s\left(\sum_{b\in\mathcal{B}_s}\mathcal{D}_b(t) - \sum_{s\in\mathcal{R}_b}\mathcal{S}_s(t)\right) \quad (20)$$

where the price in the next iteration is the difference between demand and supply at time $t$, weighted by the learning rate $\sigma$ and added to the price in the current iteration. A positive value of $\sum_b \mathcal{D}_b(t) - \sum_s \mathcal{S}_s(t)$ indicates that there is more demand, i.e., there is a shortage of data volume, and thus the price increases. Alternatively, if supply is greater than demand, the price will decrease. This process is repeated until the price difference $|\pi(t+1)-\pi(t)|$ is less than a threshold. The stability of the solution depends on the learning rate $\sigma_s$, which is analyzed in the sequel. This process is repeated for every buyer $b$ giving rise to the optimum price $\pi_{bs}^*$ and the matching of buyers and sellers $\mathbf{T}^*$.

For the practical implementation of the algorithm, communication between buyers and sellers is required only for trading and this can be done through the common control channel. The operator supervises the trading and ensures that the transactions are reflected in the buyers' and sellers' monthly bills. This will be done automatically by software in the terminals.

**Theorem 1.** The proposed Algorithm 1 is guaranteed to reach a stable matching and a data trading price.

*Proof.* Given the limited transmission range and available data at the seller, the number of alternatives for both buyers and sellers is finite. Besides, only swaps that improve all players' utility will occur. Once the stable matching is formed, the pricing algorithm converges to a stable price as in [16]. The stability of the pricing algorithm depends mainly on the learning rate. The most common way to analyze stability is to consider the eigenvalues of the Jacobian matrix of the pricing function in (20). Following [16], the fixed point $\pi_{bs}$ is stable if and only if,

$$0 < \sigma_s < \left(\frac{1}{\alpha_b\theta_b}\left(\frac{\theta_b}{\pi_{bs}}\right)^{1+1/\alpha_b} + \frac{1}{\alpha_s\theta_s}\left(\frac{\theta_s}{\pi_{bs}}\right)^{1+1/\alpha_s}\right)^{-1} \quad (21)$$

where $(\theta_b, \alpha_b)$ and $(\theta_s, \alpha_s)$ denote the parameters for buyers and sellers in the utility $f(\cdot)$.

---

**Algorithm 1** Dynamic Data Trading

1: Initialize the price $\pi_{bs}(0)$
**Stage 1 – Formation of buyer-seller associations**
2: Each buyer $b$ chooses a set of sellers $\mathcal{R}_b$ following $\succ_b$ as in (16):
  $\mathcal{R}_b = \{\tilde{U}_{b1} \geq ... \geq \tilde{U}_{bk}\}$
3: Each seller $s$ selects a set of buyers $\mathcal{B}_s$ following $\succ_s$ as in (17)
4: **repeat**
5:   Obtain $\tilde{U}_b(\eta)$ and $\tilde{U}_s(\eta)$ for the current matching $\eta$ and update $\mathcal{R}_b$
6:   **if** $(s',\eta_{ss'}^b) \succ_b (s,\eta)$ **then**
7:     Buyer $b$ sends a proposal to seller $s'$
8:     Seller $s'$ computes $U_{s'b}(\eta_{ss'}^b)$ for the swap matching $\eta_{ss'}^b$
9:     **if** $(b,\eta_{ss'}^b) \succ_{s'} (b,\eta)$ **then**
10:       $\mathcal{B}_{s'} \leftarrow \mathcal{B}_{s'} \cup \{b\}$ ;
11:       $\eta \leftarrow \eta_{ss'}^b$
14:   end
15: end
16: **Stage 2 – Data trading price**
17: **for** each buyer $b$ in the current matching:
18:   **for** each $s \in \mathcal{R}_b$:
19:     Obtain marginal data demand (18) and supply (19)
20:     Calculate learning rate $\sigma_s$ using (21)
21:     Obtain price $\pi_{bs}(t+1)$ using (20)
22:     **while** $|\pi_{bs}(t+1) - \pi_{bs}(t)| > \varepsilon$ **do**
23:       Update data demand (18) and supply (19)
24:       $t = t+1$
25:       Calculate learning rate $\sigma_s$ using (21)
26:       Update price $\pi_{bs}(t+1)$ using (20)
27:     end
28:   end
29: end
30: **until** $\nexists \eta_{ss'}^b : (s',\eta_{ss'}^b) \succ_b (s,\eta)$ and $(b,\eta_{ss'}^b) \succ_{s'} (b,\eta)$
31: Select seller $s$ for data trading at price $\pi_{bs}^*$
32: Optimum trading matrix $\mathbf{T}^*$ is obtained

---

Table 1. Price benchmark for UPN service

|  | $\psi_{b0}=0, \psi_{bs}=0$ | $\psi_{b0}=1, \psi_{bs}=0$ | $\psi_{b0}=0, \psi_{bs}=1$ | $\psi_{b0}=1, \psi_{bs}=1$ |
|---|---|---|---|---|
| $e_b=1$ | $\pi < p$ | $\pi < p - \kappa_0/Q_b$ | $\pi < p - \kappa_s/Q_b$ | $\pi < p - \chi/Q_b - \kappa_0/Q_b$ |
| $1-e_b=1$ | $\pi < (f(Q_b) + rQ_b - f(Q_b^i))/Q_b$ | $\pi < (f(Q_b) + rQ_b - f(Q_b^i))/Q_b - \kappa_0/Q_b$ | $\pi < (f(Q_b) + rQ_b - f(Q_b^i))/Q_b - \kappa_s/Q_b$ | $\pi < (f(Q_b) + rQ_b - f(Q_b^i))/Q_b - \kappa_0/Q_b - \kappa_s/Q_b$ |

### C. Price Benchmark for UPN service

Following the model from Section III, we derive the price conditions under which a service-swap will occur. Let us assume that the user requests the same amount of data in the macro-cell and the UPN, $Q_b$. The user will prefer to transmit in the UPN if $U_{bs} > U_{b0}$, which occurs when

$$\pi < (1-e_b)\frac{\left[f(Q_b) + rQ_b - f(Q_b^i)\right]}{Q_b} + e_b p - \frac{\kappa_0\psi_{b0}+\kappa_s\psi_{bs}}{Q_b} \quad (22)$$

where $\kappa_0$ and $\kappa_s$ are the penalty factor for violating the connectivity constraint in the macro-cell and the UPN, $\psi_{b0}$ and $\psi_{bs}$, respectively, as explained in Section IV.A. Depending on the service and data availability, different price benchmarks are obtained for the UPN service, as shown in Table 1.

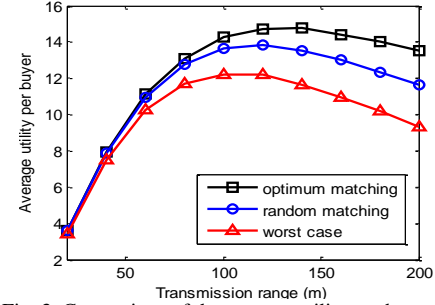
Fig. 3. Comparison of the average utility per buyer.
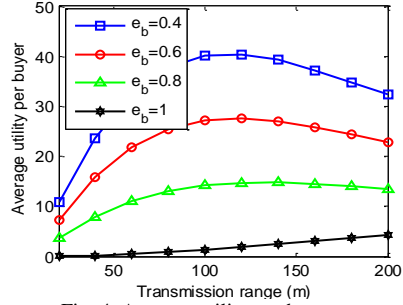
Fig. 4. Average utility per buyer.
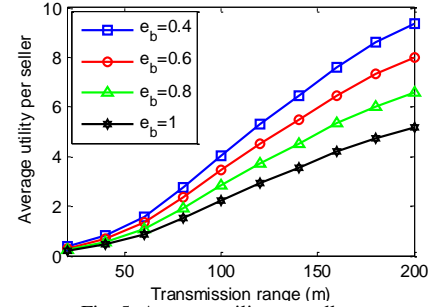
Fig. 5. Average utility per seller.

If the user exceeds the data plan ($e_b = 1$) and both the macro-cell and UPN service satisfy the required QoS, the user will choose UPN if $\pi < p$. If the macro-cell service does not provide the required QoS, the benchmark price to connect to UPN is $\pi < p - \kappa_0/Q_b$, where there is a penalty discount. The same penalty is assumed for the UPN. Similarly, if the UPN does not provide the required QoS, the price is $\pi < p - \kappa_s/Q_b$ where $\chi$ is the penalty in the UPN. The same analysis can be applied when the user does not exceed the data plan ($1-e_b=1$).

## V. SIMULATION RESULTS AND ANALYSIS

We consider a single macro-cell network deployed in a square area of 1000 x 1000 $m$ with the BS at the center. The buyers and sellers are randomly placed in the network. We set all users' transmit power to 20 mW, the noise level to $\sigma^2 = 10^{-7}$ W/Hz and the propagation loss to $\alpha = 3$. We assume that buyers and sellers have a monthly contract for a data volume of 10 GB. The minimum SINR requirement varies between [1, 20] dB and the duration of connectivity varies between [0, 15] minutes. The energy consumption $\xi$ is set to 0.257 J/MB as estimated in [17] for 4G connection. The data usage price depends on the country, the data plan and the service provider. Based on a recent ITU publication [18], we set the price of a 10GB data plan to 10€, and the price per GB traded in UPN between 0.1 and 1€. The results are averaged over 1000 runs.

Figure 3 shows the average utility per buyer obtained for a network with 10 sellers and 20 buyers, where the probability of exceeding the data plan is $e_b=0.8$. The results are presented with respect to the transmission range, which varies from 20 to 200 m; this corresponds to a minimum SINR of 20 dB to 1dB in the same range. The performance is compared with random matching and worst case utility. The latter refers to the matching that provides the lowest buyer´s utility. Note that these schemes were selected for comparison purposes since, to the best of our knowledge, this is the first paper to address the association problem for data trading in UPNs. Fig. 3 shows that, as the transmission range increases and with it the options to connect, our proposed scheme yields a performance advantage of 25% in terms of utility improvement relative to random match and 50% to worst case utility. Similar gains were observed in seller utility but the results are omitted due to space limitations. Buyer utility, by contrast, decreases with transmission range as the SINR is lower.

Figures 4 and 5 present the average utility per buyer and seller, respectively, for different values of the probability of exceeding the data plan, $e_b$. As before, we assume 10 sellers and 20 buyers. We can see that, as $e_b$ increases, the utility for buyers and sellers decreases. This is because buyers need to pay extra for the additional data exceeding the data plan. Therefore, the amount of data requested is significantly lower than the initial data plan. The average volume of data traded per UPN in this scenario varied between 0.5 and 2 GB. It is worth noting that seller utility increases with transmission range as the number of potential buyers increases.

Figure 6 shows the average number of buyers transmitting to the macro-cell BS, as the number of buyers $B$ increases and the number of sellers remains constant to $S = 10$. For high $e_b$ values, the number of users transmitting to the BS decreases and buyers prefer to join the UPN to purchase additional data. Furthermore, if price $p$ in the macro-cell increases with respect to the price in the UPN $\pi$, an additional decrease in number of users is observed in the macro-cell. In particular, when price $p$ doubles, the number of users transmitting to the BS decreases by about 25%.

Figure 7 shows the revenue of the operator obtained in the UPN and BS operation. We can see that the UPN yields higher revenue to the operator than the BS as a higher volume of data is transmitted through the UPN. The revenue reaches 200% when users have high QoS requirements (low transmission range).

Figure 8 shows the number of iterations needed for convergence of the data trading algorithm in stages 1 and 2 as $B$ varies. Note how the number of iterations increases for medium-size networks, $B \leq 50$, and remains constant for $B > 50$. This is because as density increases, adjacent sellers will offer similar performance levels and thus, the number of swaps will decrease. It is worth noting that the convergence of stage 2 is influenced by price initialization. If the price is initialized to the global market price, which is obtained as in (20) considering the total data demand and supply in the network, the number of iterations is significantly reduced. This price could be provided by the operator as a reference price for data trading. Still, the algorithm presents a reasonable convergence time even in very dense networks.

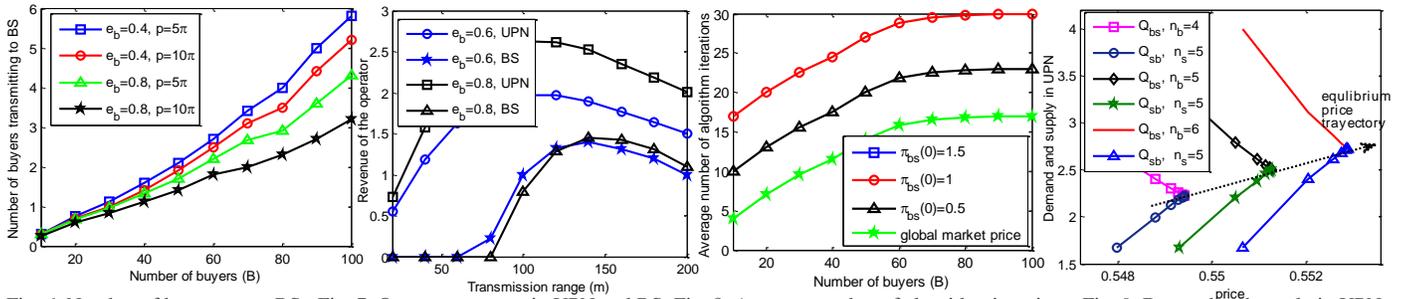

Fig. 6. Number of buyers tx. to BS; Fig. 7. Operator revenue in UPN and BS; Fig. 8. Average number of algorithm iterations; Fig. 9. Demand and supply in UPN.

The equilibrium price in the UPN is presented in Figure 9 as a function of data demand and supply. Each pair of curves represents the demand and supply in a UPN consisting of $n_b$ and $n_s$ potential buyers and sellers, respectively. The considered pricing scheme based on market equilibrium adjusts the price based on the data demand and supply. The equilibrium price is obtained when these are equal. When the demand, $n_b$, increases for a constant supply, $n_s$, the price increases accordingly. This pricing scheme controls and incentivizes the demand for data in the UPN.

## VI. Conclusions

In this paper, we have proposed an analytical market model for data trading in a UPN that captures the preferences and connectivity requirements of buyers, sellers and macro-cell BSs. We have formulated the buyer-seller association problem for data trading as a matching game in which buyers, sellers and BSs rank each other according to their preferences, and in which, the trading price is governed by market equilibrium. In this game, preferences are interdependent and influenced by matchings that arise. To solve the game, we have developed a distributed algorithm that accounts for network externalities. We have shown that, with the proposed algorithm, buyers, sellers and BSs reach a stable matching in a reasonable number of iterations. Our simulation results showed that the proposed approach can provide significant gains in terms of utility, with gains of up to 25% observed relative to random matching and up to 50% relative to worst case utility. Significant gains were also observed for the operator. We also analyzed the properties of the stable associations.